\documentclass[aps,prc,showpacs,preprintnumbers,superscriptaddress,nofootinbib,f
loatfix,11pt]{revtex4-1}
\usepackage{amsfonts,amssymb,stmaryrd,latexsym,amsmath}
\usepackage{subfigure,graphicx}
\usepackage{times}
\usepackage{slashed}
\usepackage[colorlinks]{hyperref}

\usepackage{graphicx}
\usepackage{amsmath, amssymb, bm}
\usepackage{dcolumn}
\usepackage{bm}
\usepackage{comment}
\usepackage{tabularx}
\usepackage{multirow}
\usepackage{braket}

\begin{document}


\title{\bf{A note on scalar meson dominance}}

\author{Y. \"Unal}
\affiliation{\textit{Helmholtz-Institut f\"ur Strahlen- und Kernphysik and Bethe Center for Theoretical Physics,
    Universit\"at Bonn, D-53115 Bonn, Germany}}
\affiliation{\textit{Physics Department, \c{C}anakkale  Onsekiz Mart University, 17100 \c{C}anakkale, Turkey}}
\author{Ulf-G. Mei{\ss}ner}
\affiliation{\textit{Helmholtz-Institut f\"ur Strahlen- und Kernphysik and Bethe Center for Theoretical Physics,
    Universit\"at Bonn, D-53115 Bonn, Germany}}
\affiliation{\textit{Institute for Advanced Simulation, Institut f\"ur Kernphysik
and J\"ulich Center for Hadron Physics, Forschungszentrum J\"ulich, D-52425 J\"ulich, Germany}}
\affiliation{\textit{Tbilisi State University, 0186 Tbilisi, Georgia}}
\date{\empty}

\begin{abstract}
We consider chiral perturbation theory with an explicit broad $\sigma$-meson and study its contribution
to the scalar form factors of the pion and the nucleon. Our goal is to learn more about resonance
saturation in the scalar sector.
\end{abstract}

\maketitle



\pagebreak

\section{Introduction}

The lowest-lying resonance in QCD is the broad $\sigma$-meson (also called $f_0(500)$), with its parameters
precisely determined from various dispersion-theoretical analyses of pion-pion scattering,
see e.g.~\cite{Caprini:2005zr,Pelaez:2015qba}.
Still, due to its large width, the  $\sigma$ (and other scalar mesons) plays a rather different role in the
low-energy effective
field theory of QCD than the vector or axial-vector mesons. The latter clearly saturate the low-energy constants to
which they can contribute, as known since long~\cite{Ecker:1988te,Donoghue:1988ed}.  In the scalar sector,
the heavier mesons like the $f_0(980)$ and $a_0(980)$ are considered in the studies of resonance saturation,
but their contribution to  the low-energy constants is small, except for $L_5$ and $L_8$, which, however,
are used to fix certain scalar couplings~\cite{Ecker:1988te}. Later, it was argued in
Ref.~\cite{Meissner:1990kz} that the imaginary part of the pion scalar form factor can be understood in
terms of a broad scalar meson, building upon the detailed investigations in \cite{Donoghue:1990xh,Gasser:1990bv}.
Further, in Ref.~\cite{Bernard:1996gq} it was shown that the dimension-two
low-energy constant $c_1$, that parametrizes the leading explicit chiral symmetry breaking in the effective
pion-nucleon Lagrangian, can be saturated by scalar meson exchange for a particular value of the ratio
$M_\sigma/g_{\sigma NN}$, with $M_\sigma$ the mass of the $\sigma$ and $g_{\sigma NN}$ the coupling of the 
$\sigma$ to nucleons. Note that in the two-nucleon system the $\sigma$-meson essentially parametrizes the
nuclear binding (in the one-boson-exchange approximation).
In fact, for some particular boson-exchange model, this ratio is close to the one required 
by resonance exchange saturation. For details, the reader is referred to Ref.~\cite{Bernard:1996gq}.

In this note, we want to reconsider the  $\sigma$-meson  contribution to the scalar form factors of the pion
and the nucleon in a chiral perturbation theory calculation at one loop, where the
effective Lagrangian is supplemented by an explicit broad scalar meson.  A comparison with the existing 
precision calculations of the these form factors will allow us to draw conclusion about the role of the
$f_0(500)$ in the context of resonance saturation. We also note that our approach is  not
the scale-chiral perturbation theory proposed in Ref.~\cite{Crewther:2013vea}, which considers
the broken conformal symmetry of QCD. The effective Lagrangian approach to that phenomenon was
originally developed in Refs.~\cite{Isham:1970xz,Schechter:1993tc}.

The paper is organized as
follows: In Sec.~\ref{sec:def} we give the basic definitions of the scalar form factors under investigation.
Then, in Sec.~\ref{sec:pion} we calculate the imaginary part of the pion scalar form factor and compare with the
precise determination based on dispersion theory. Sec.~\ref{sec:sigma} contains the analogous calculation
of the nucleon scalar form factor and the comparison with the corresponding nucleon spectral function at two loops
in heavy baryon chiral perturbation theory.
We end with a short summary.

\vfill

\section{Scalar form factors}
\label{sec:def} 
The scalar form factor of the pion and the nucleon are defined via the matrix element of the
scalar quark density $\bar{q}q$ in the isospin symmetry limit
\begin{equation}
\begin{aligned}
  F_{\pi}^{S}(t)\;&=\;\Braket{\pi(p_{f})|\hat{m}(\bar{u}u+\bar{d}d)|\pi(p_i)}~,\\
  \sigma(t)\;&=\;\frac{1}{2m_N}\Braket{N(p_{f})|\hat{m}(\bar{u}u+\bar{d}d)|N(p_i)},
\label{ff}
\end{aligned}  
\end{equation}
with $t=(p_f-p_i)^2$  the invariant momentum transfer squared, $m_N$ the nucleon
mass and $\hat{m}=(m_u+m_d)/2$ the average light quark mass. 
The scalar form factor of the nucleon satisfies the once-subtracted dispersion relation
\begin{equation}
\sigma(t)\;=\;\sigma_{\pi N}+\frac{t}{\pi} \int_{t_0}^{\infty} dt'\frac{\text{Im} \sigma(t')}{t'(t'-t)}
\end{equation}
where $t_0$ is the threshold energy for hadronic intermediate states. The $\pi N$ $\sigma$-term
$\sigma_{\pi N}=\sigma(0)$ is defined via the Feynman-Hellman theorem at $t=0$, similar to the form factor
$F_\pi^S$,  which gives the pion $\sigma$-term
at zero momentum transfer. It is known that $F_\pi^S (0) = (0.99\pm 0.02)M_\pi^2$~\cite{Gasser:1990bv},
so in what follows we consider the normalized scalar pion form factor  $F_\pi^S (t)/M_\pi^2$. Here, $M_\pi$
denotes the charged pion mass. In this paper, we want to investigate the $\sigma$-meson
contribution to these scalar form factors and draw some conclusion on the related issue of scalar meson dominance.

\section{The  scalar form factor of the pion}
\label{sec:pion}

\begin{figure}[t!]
\centering
\includegraphics[width=1\textwidth]{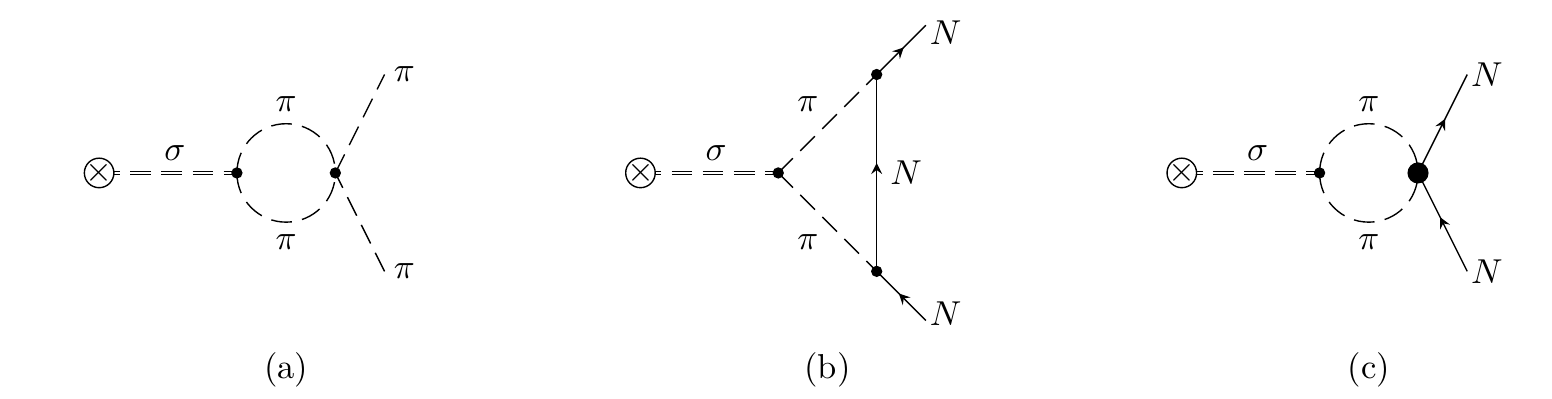}
\caption{$\sigma$-meson contribution to the 
  isoscalar scalar form factors of the pion (a) and the nucleon (b,c). Dashed and solid lines
  denote pions and nucleons, respectively. The double dashed lines and the cross represent the $\sigma$-meson and
  the scalar source, respectively. The light dots represent leading order vertices while the  heavy dot characterizes a 
  dimension-two pion-nucleon  vertex.}
\label{fig:sigsca}
\end{figure}

In this section, we want to calculate the imaginary part of the scalar pion form factor
with a particular emphasis on the contribution from the broad $\sigma$-meson, cf. Fig.~\ref{fig:sigsca}(a).
For definiteness, we use
\begin{equation}
M_\sigma= 478\,{\rm MeV}~,~~~~ \Gamma_\sigma= 324\, {\rm MeV}~~.
\end{equation}  
Further, we adopt the choice in \cite{Deandrea:2000ce} for $\sigma$-propagator, 
\begin{eqnarray}
S_\sigma(t)&=&\frac{1}{t-M_\sigma^2+i\Gamma_\sigma(t)M_\sigma} \nonumber  
\end{eqnarray}
where the co-moving width $\Gamma_\sigma(t)$ is given by
\begin{equation}
\Gamma_\sigma(t)=\Gamma_{\sigma}\frac{M_\sigma}{\sqrt{t}}\frac{\sqrt{t/4-M_\pi^2}}{\sqrt{M_\sigma^2/4-M_\pi^2}}
\end{equation}
with momentum transfer squared $t$, see also the discussion in Ref.~\cite{Gardner:2001gc}. 

The following power counting rules for the loop diagrams are used (we consider here the pion and the nucleon case
together to keep the presentation compact):
vertices from $\mathcal{L}^{(n)}$ count as $\mathcal{O}(q^n)$, so we count the nucleon propagator
as $\mathcal{O}(q^{-1})$, and the sigma and pion propagators as
$\mathcal{O}(q^{-2})$. Thus the chiral order of the diagram in Fig.~\ref{fig:sigsca}(a) is $\mathcal{O}(q^3)$,
as only lowest order vertices from $\mathcal{L}^{(2)}_{\pi\pi}$ are involved, and the diagrams in
Fig.~\ref{fig:sigsca}(b)-(c) are $\mathcal{O}(q^4)$ at low energies, i.e. for small $t$ (for precise
definitions of the pion-nucleon Lagrangian, see Sec.~\ref{sec:sigma}).

Let us briefly discuss the kinematics for evaluating the diagrams shown in Fig.~\ref{fig:sigsca}.
We start with the nucleon case, as it is more general.
We work in the center-of-momentum frame of the nucleon pair with 
$q=\big(-2E_p,~\vec{0}\,\big)$. The initial and the final momentum of the nucleons are, respectively, 
$p_i=\big(E_p,~\vec{p}\,\big)$, $p_f=\big(-E_p,~\vec{p}\,\big)$, with $|\vec{p}\,|= (t/4 - m_N^2)^{1/2}$ and 
$E_p = (m_N^2+|\vec{p}\,|^2)^{1/2}$.
The imaginary part of the loop diagram corresponds to a cut diagram for
the momentum transfer squared $t \geq 4M_{\pi}^2$. For this calculation, we perform a reduction
to scalar loop integrals and thus require the basic scalar loop integrals of one- and two- point functions, respectively,

\begin{equation}
\begin{aligned}
A_0(m^2)=& \frac{(2 \pi \mu)^{4-n}}{i \pi^2}\int \frac{d^nk}{k^2-{m}^2+i\epsilon^+},   \\
B_0(q^2, m_1^2, m_2^2)=& \frac{(2 \pi \mu)^{4-n}}{i \pi^2} \int \frac{d^nk}{[k^2-m_1^2
       +i\epsilon^+][(k+q)^2-m_2^2+i\epsilon^+]} 
\end{aligned}
\label{Int}
\end{equation}
with $q^2=t=(p_f-p_i)$ and $\epsilon^+$ stands for $\epsilon \to 0^+$.  For the pion case,
we need to consider these functions with $m=m_1=m_2=M_\pi$ and the corresponding kinematical
variables are simply obtained by the substitution $m_N \to M_\pi$ in the above-mentioned
formulas.

The one-loop contribution depicted in Fig.~\ref{fig:sigsca}(a) is readily evaluated
\begin{equation}
\label{eq:ImPion}
\begin{aligned}
  \text{Im}~F_{\pi}^S(t)\;=\;&\frac{B g_\sigma g_{\sigma \pi \pi} \Big(A_0(M_{\pi}^2)(12 t-14 M_{\pi}^2)+(15M_{\pi}^2
    t-6M_{\pi}^4-6t^2 )\, \text{Re}[{B_0(t, M_{\pi}^2, M_{\pi}^2)}]\Big)}{3F_{\pi}^4 \pi^2}\\
&\times \frac{-M_{\sigma} \Gamma_{\sigma}(t) }{t^2+M_{\sigma}^4-2 M_{\sigma}^2 t + M_{\sigma}^2 \Gamma_{\sigma}^2 (t) }
\end{aligned}
\end{equation}
where
\begin{equation}
\begin{aligned}
A_0(m^2)=\; & -2 m^2 \log\Big(\frac{m}{\mu}\Big), \\ 
B_0(t,m_1^2,m_2^2)=\; &1-2 \log\Big(\frac{m_{1}}{{\mu}}\Big)-\frac{(t-m_1^2+m_2^2)\log\Big(\frac{m_{2}}{{m_1}}\Big)}{t}   \\
& - \frac{2m_1 m_2 \sqrt{1-\frac{(m_1^2+m_2^2-t)^2}{4m_1^2 m_2^2}} \text{arccos}
  \Big[\frac{m_1^2+m_2^2-t}{2 m_1 m_2}\Big]}{t}. \nonumber
\end{aligned}
\end{equation}

\begin{figure}[t!]
\centering
\includegraphics[width=0.65\textwidth]{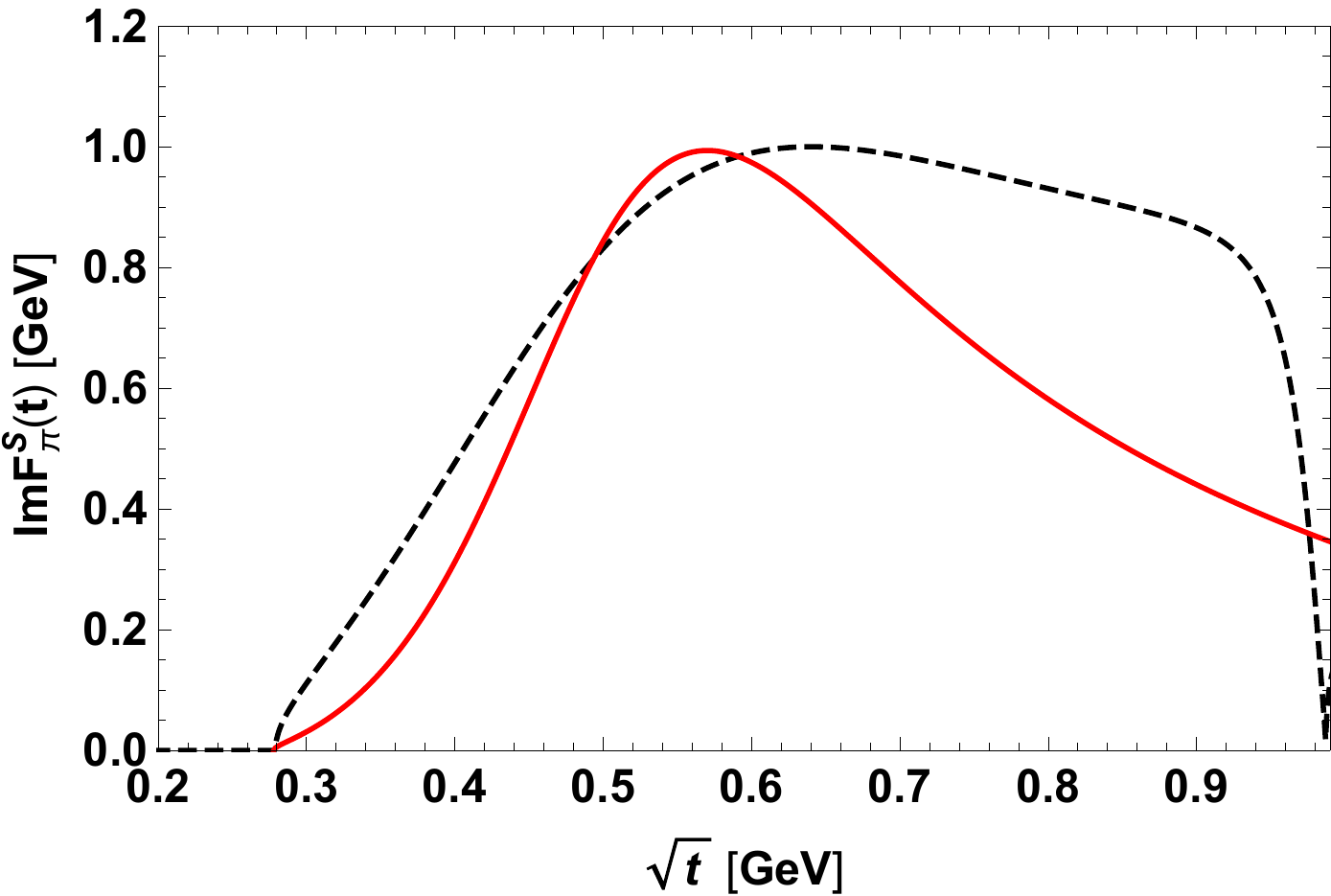}
\caption{Imaginary part of the scalar pion form factor based on Eq.~(\ref{eq:ImPion}) (solid red line)
  in comparison to the dispersion-theoretical result of Ref.~\cite{Ropertz:2018stk} (dashed black line).}
\label{fig:ImPFF}
\end{figure} 

The following values for the various masses and couplings constants are:
$M_\pi = 0.139$~GeV, $F_\pi = 0.092$~GeV, and $g_{\sigma \pi \pi} = 2.52\,\text{GeV}$
from the experimental value of $\sigma$ width. 
Further, we use $B = 0.7$~GeV, but note that its precise value depends on the choice of the
average quark mass. The coupling $g_\sigma$ is merely used for normalization.

In Fig.~\ref{fig:ImPFF} we show the imaginary part of the scalar pion form factor in our
approach in comparison to the dispersion-theoretical analysis of Ref.~\cite{Ropertz:2018stk}. Up to
$\sqrt{t} \simeq 0.6\,$GeV, the curves are very similar, but of course the $f_0(980)$
contribution that causes the dip at $\sqrt{t} \simeq 1\,$GeV is not captured in our
approach. Still, the visible enhancement due to the broad $\sigma$ is clearly reflected
in the imaginary part.

\section{The  scalar form factor of the nucleon}
\label{sec:sigma}

In this section we focus on the calculation of the imaginary part of the isoscalar nucleon
scalar form factor generated from the $\pi\pi$ intermediate states based on relativistic
two-flavor baryon chiral perturbation theory. At lowest order in the quark mass and momentum expansion,
the relevant interaction Lagrangians are given by \cite{Fettes:2000gb, Gasser:1987rb,Ecker:1988te}
\begin{equation}
\begin{aligned}
\mathcal{L}^{(1)}_{\pi N}=
&\;   \frac{g_A}{2}\bar{\Psi}\gamma^{\mu}\gamma_{5}u_{\mu}\Psi,   \\
\mathcal{L}^{(2)}_{\pi N}=
&\;   c_1\langle\chi_{+}\rangle\bar{\Psi}\Psi-\frac{c_2}{4m_{N}^2}\langle u_{\mu}u_{\nu}\rangle (\bar{\Psi}D^{\mu}D^{\nu}
\Psi+h.c)   \\
&\;  +\frac{c_3}{2}\langle u_{\mu}u^{\mu}\rangle \bar{\Psi} \Psi-\frac{c_4}{4}\bar{\Psi}\gamma_{\mu}
\gamma_{\nu}[u^{\mu},u^{\nu}]\Psi + ...,   \\
\mathcal{L}^{(2)}_{\sigma \pi \pi}=
&\;  g_{\sigma \pi \pi} \sigma \langle{u_\mu u^\mu}\rangle \,\, .
\end{aligned}
\label{L}
\end{equation}
Here, $\Psi$ denotes the nucleon doublet, $\chi_{+}=u^\dagger \chi u^\dagger+u\chi^{\dagger}u$,  with $\chi=2B_0(\mathcal{M}+ s)$
where $s$ represents the external scalar source and $\langle \ldots \rangle$ denotes
the trace in flavor space. Further, $g_A$ is the nucleon axial-vector coupling, $g_A \simeq 1.27$, and the
low-energy constants  $c_2, c_3$ and $c_4$ are taken as $c_{2} = 3.13 m_N^{-1}$, $c_{3} = -5.61 m_N^{-1}$ and
$c_{4} = 4.26 m_N^{-1}$\cite{Hoferichter:2015tha}. These LECs are not affected by a contribution
from the $\sigma$, see Ref.~\cite{Bernard:1996gq}. As already discussed in the introduction,
the $\sigma$-meson contributes to the LEC $c_{1}$. We consider two extreme cases, namely $c_{1} =0$, which
corresponds to a complete saturation of this LEC by the light scalar meson and $c_{1} =0.55  m_N^{-1}$,
which is half of the value given in Ref.~\cite{Hoferichter:2015tha}. This latter scenario leaves
room for other contributions to this particular LEC.

In addition to the scalar loop integrals in Eq.~(\ref{Int}) we also need the integral of three-point function as
\begin{equation}
\begin{aligned}
&C_0(p_i^2, (p_f-p_i)^2, p_f^2, m_1^2, m_2^2, m_3^2)=   \\
&\qquad\qquad \frac{(2 \pi \mu)^{4-n}}{i \pi^2}\int \frac{d^nk}{[k^2-m_1^2
       +i\epsilon^+][(k-p_i)^2-m_2^2+i\epsilon^+][(k-p_f)^2-m_3^2+i\epsilon^+]}
\end{aligned}
\label{Int3}
\end{equation}
with $q^2=t=(p_f-p_i)$ and $\epsilon^+$ stands for $\epsilon \to 0^+$. 
From these, the expressions for the imaginary part of the scalar
form factor, which has dimension [mass], is given as 
\begin{equation}
\begin{aligned}
\text{Im}~\sigma_{N}(t)\;=\;&\frac{B g_\sigma g_{\sigma \pi \pi}}{6 F_{\pi}^4 {\pi}^2 m_N} \Bigg(\frac{18 g_A^2 m_{N}^3}{4m_{N}^2-t}\Big((8m_{N}^2-2t)A_0(m_{N}^2)+(2M_{\pi}^2-t)(4m_N^2 M_\pi^2-2 m_N^2 t)\\
&\times \text{Re}[{C_0(m_N^2, t, m_N^2, m_N^2, M_{\pi}^2, M_{\pi}^2)}]-t (2M_{\pi}^2-t)\, \text{Re}[{B_0(t, M_{\pi}^2, M_{\pi}^2)}]\\
&+(16m_{N}^2 M_{\pi}^2-4m_N^2 t-2M_{\pi}^2 t) B_0(m_N^2, m_N^2,M_{\pi}^2)\Big)\\
&- 6 \Big (2 m_N^2 M_{\pi}^2(24 c_1-5c_2-24c_3)+2t(m_N^2 c_2-M_{\pi}^2 c_2\\
&+6 m_N^2 c_3)+c_2 t^2 \Big) A_0(M_{\pi}^2)-\Big(8m_N^2 M_{\pi}^2(6c_1-c_2-3c_3)\\
&+2t(m_N^2 c_2+M_{\pi}^2c_2+6m_N^2c_3)+c_2 t^2\Big)(6M_{\pi}^2-3 t)\,\text{Re}[{B_0(t, M_{\pi}^2, M_{\pi}^2)}]\\
&+c_2\Big(66 m_N^2 M_\pi^4 + 4 m_N^2 t^2+8 M_{\pi}^2 t^2-32m_N^2 M_\pi^2 t-12 M_{\pi}^4 t-t^3 \Big)\Bigg)\\
& \times \frac{-M_{\sigma} \Gamma_{\sigma}(t) }{t^2+M_{\sigma}^4-2 M_{\sigma}^2 t + M_{\sigma}^2 \Gamma_{\sigma}^2 (t)} 
\end{aligned}
\end{equation}

\begin{figure}[t!]
\includegraphics[width=.65\textwidth]{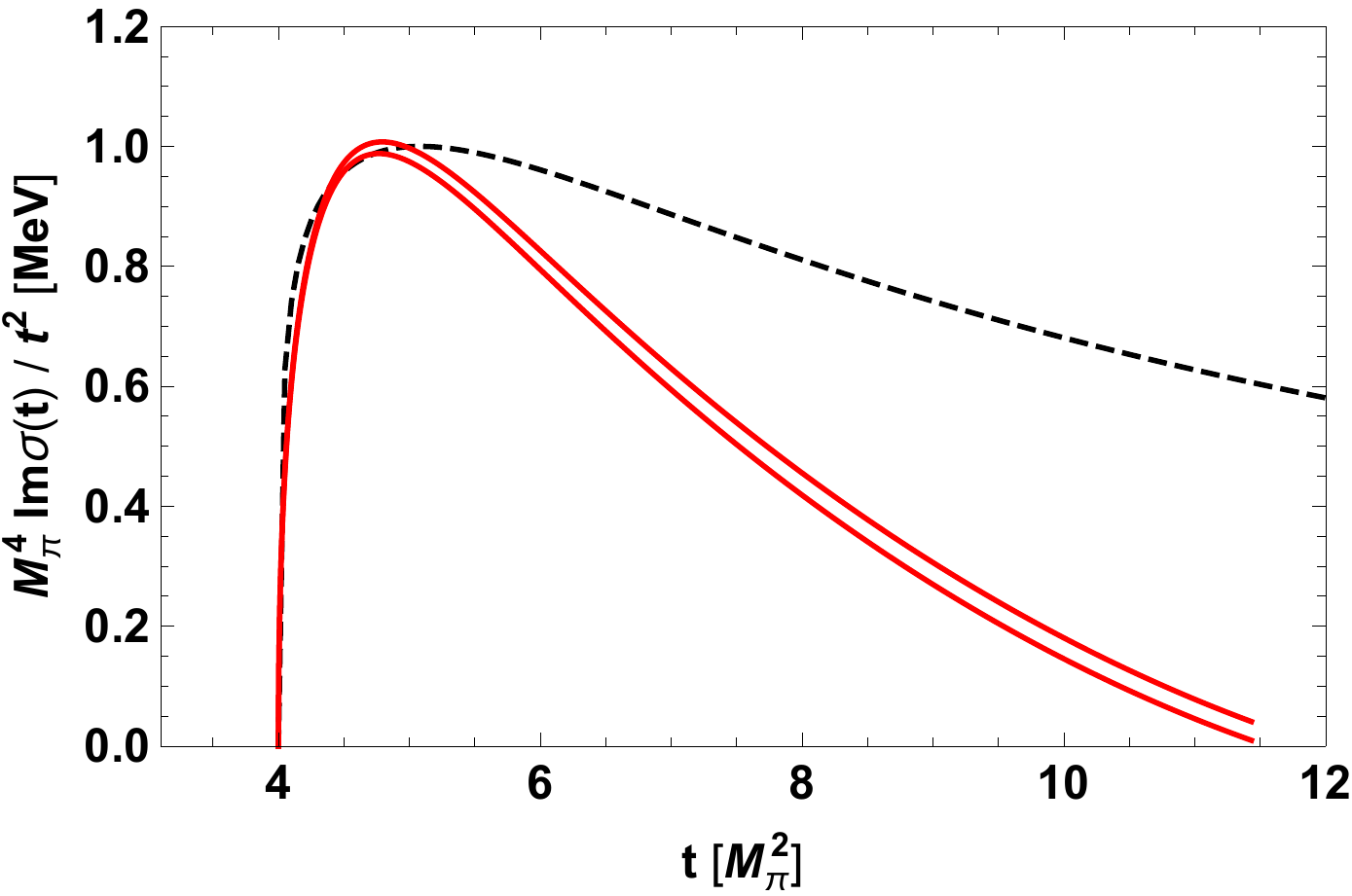}
\centering
\caption{$\sigma$-meson contribution (solid lines) to the isoscalar spectral function of the scalar nucleon form factor
  multiplied with $M_\pi^4/t^2$,  compared  with the two-loop chiral perturbation theory result of
  Ref.~\cite{Kaiser:2003qp}  (black dashed line). The lower (upper) solid line refers to the case of complete (partial) saturation
  of the LEC $c_1$ as discussed in the text.}
  \label{fig:specsigN}
\end{figure} 
\noindent
where the $A_0(m^2)$ and $B_0(m_1^2, m_1^2, m_2^2)$ functions are real. 

The resulting weighted spectral function $M_\pi^4 \,{\rm Im}~\sigma(t)/t^2$ is shown in Fig.~\ref{fig:specsigN} for the
case of complete saturation $c_1=0$ (lower solid red line) and the one of partial saturation ($c_1 = -0.55 m_N^{-1}$) (upper
solid red line) in comparison to the two-loop heavy baryon chiral perturbation theory results of Ref.~\cite{Kaiser:2003qp}.
We see that the explicit $\sigma$-meson contribution drops faster than the pion loop contribution, showing that the
$\sigma$ does not saturate this imaginary part. 

\section{Summary}

In this note, we have considered the broad $\sigma$-meson contribution to the scalar form factors of the
pion and the nucleon, respectively. In the pion case, the imaginary  part clearly exhibits the $f_0(500)$ contribution,
but below $\sqrt{t} \simeq 1$~GeV, one also needs to include the $f_0(980)$. The latter is responsible for the
pronounced dip in the imaginary part. Concerning  resonance saturation, just including the scalar mesons
around 1~GeV is not sufficient, though one can produce the light scalar as a rescattering effect through pion loop resummation.
This, however, requires a non-perturbative framework. Similarly, for the scalar nucleon form factor, we find that
the $\sigma$-meson saturates the imaginary part at low invariant momenta but drops faster than the two-loop
contribution. This is similar to the findings in Ref.~\cite{Bernard:1996gq}, where it was shown that the leading
scalar-isoscalar low-energy constant $c_1$ can only be explained in terms of scalar meson exchange for a very
special combination of ratio of the sigma-nucleon coupling constant to the $\sigma$ mass. As the calculations presented
here underline, the broad $\sigma$-meson enjoys a very special role in low-energy QCD.

\section*{Acknowledgments}
We thank Norbert Kaiser and Bastian Kubis for providing us with their results.
We are also grateful for the referee for a pertinent comment.
Partial financial support from the Deutsche Forschungsgemeinschaft (SFB/TRR~110, ``Symmetries
and the Emergence of Structure in QCD''), by the Chinese 
Academy of Sciences (CAS) President's International Fellowship Initiative (PIFI) (grant no. 2018DM0034) 
and by VolkswagenStiftung (grant no. 93562) is acknowledged.

\end{document}